\documentclass[a4paper]{jpconf}
\usepackage{graphicx}
\usepackage{amsmath,bm}
\usepackage{verbatim}
\usepackage{amsthm,mathtools}

\newcommand{\be}{\begin{equation}}
\newcommand{\ee}{\end{equation}}

\begin{document}
\title{Thermodynamics at zero temperature: \\inequalities for the ground state \\ of a quantum many-body system}

\author{N Il'in$^{1}$, E Shpagina$^{2,3}$ ,O Lychkovskiy$^{1,4,5}$}

\address{$^1$ Skolkovo Institute of Science and Technology,	Bolshoy Boulevard 30, Moscow 121205, Russia}
\address{$^2$ Faculty of Physics, NRU Higher School of Economics, 101000, Myasnitskaya 20, Moscow, Russia}
\address{$^3$ Institute of Solid State Physics, Russian Academy of Sciences, 142432 Chernogolovka, Russia}
\address{$^4$ Department of Mathematical Methods for Quantum Technologies, Steklov Mathematical Institute of Russian Academy of Sciences,	Gubkina str., 8, Moscow 119991, Russia}
\address{$^5$ Laboratory for the Physics of Complex Quantum Systems, Moscow Institute of Physics and Technology, Institutsky per. 9, Dolgoprudny, Moscow  region,  141700, Russia}

\ead{o.lychkovskiy@skoltech.ru}

\begin{abstract} We prove that for a single-component many-body system at zero temperature the inequality $E_{\rm int} \leq\,  P\,V$ holds, where $E_{\rm int}$ is the interaction energy, $P$ is pressure and $V$ is volume. This inequality is proven under rather general assumptions with the use of Anderson-type bound relating ground state energies of systems with different numbers of particles. We also consider adding impurity particles to the system and derive inequalities on the chemical potential of the impurity and binding energy of the bound state of two impurities.
\end{abstract}

\section{Introduction} Inequalities relating thermodynamic quantities play a profound role in thermodynamics and statistical physics of quantum many-body systems. Examples include rigorous proofs of the second law of thermodynamics \cite{Allahverdyan,skrzypczyk2014work}, conditions on superfluidity \cite{andreev2003thermodynamic,Lychkovskiy}, inequalities involving correlation functions in the vicinity of a critical point \cite{Griffiths}, inequalities on the derivative of the free energy with respect to the scattering
length in quantum gases \cite{Werner_I,Werner_II}, proofs of the existence of the long range order in spin systems \cite{nishimori1989ground,ozeki1989long} {\it etc.} Here we address a quantum system of particles with a  pairwise interaction in its ground state (i.e. ``at zero temperature''). For a single-component system, we derive inequalities relating its energy per particle, $\varepsilon(\rho,m,g)$ and its derivatives. We also consider a system with added particles of a different sort (impurities) and derive inequalities on the chemical potentials and binding energies of the impurities. The derivation of the inequalities is based on Anderson's general idea relating the ground state energy of a larger system to the ground state energies of several smaller systems \cite{Anderson}. We illustrate our results by applying them to various cases where the ground state energy is known approximately or exactly.

\section{Single species}

\medskip

\subsection{Preliminaries}

\medskip

Consider a system of $N$ particles in a volume $V$. Its Hamiltonian in the first-quantized form reads
\begin{equation}\label{H homogeneous}
H=\sum_{1\leq i \leq N} \frac{\bm{p}_i^2}{2m} + g \sum_{1\leq i<j \leq N} U(\bm{r}_i-\bm{r}_j).
\end{equation}
Here $m>0$ is the mass of a particle, $\bm{p}_i$ and $\bm{r}_i$ are, respectively, the operators of the momentum and the coordinate of the $i$'th particle and $g\, U(\bm{r}_i-\bm{r}_j)$ is the interparticle interaction potential. We single out a multiplicative constant $g$ to be able to take partial derivatives with respect to it, and assume $g>0$ (the sign of $U(\bm{r}_i-\bm{r}_j)$ can be arbitrary and can depend on the  argument $(\bm{r}_i-\bm{r}_j)$). We do not specify the statistics of the particles.  Also the dimensionality $D=1,2,3$ of the system is not fixed.

The ground state energy of the system reads
\begin{equation}\label{E_N}
E_N(V,m,g) = \inf_{\Psi:\, \|\Psi\|=1} \langle \Psi | H |  \Psi \rangle,
\end{equation}
where infimum  is over all normalized states $\Psi$ in the appropriate Hilbert space. We have equipped $E_N(V,m,g)$ with the arguments indicating the number of particles, the volume and the parameters of the Hamiltonian for further use.

We assume that the ground state energy can be written as
\begin{equation}\label{varepsilon}
E_N(V,m,g) = N\, \big(\varepsilon(\rho,m,g) + \delta_N(\rho,m,g)\big)
\end{equation}
in the thermodynamic limit of $N,V\rightarrow\infty$ with the particle density $N/V=\rho$ kept constant, where  $\varepsilon(\rho,m,g)$ is the energy per particle, and $\delta_N(\rho,m,g)$ is the boundary contribution to the energy with the property
\begin{equation}\label{boundary scaling 1}
\partial_\rho \delta_N(\rho,m,g)=o(1),\quad \partial_g \delta_N(\rho,m,g)=o(1)
\end{equation}
in the thermodynamic limit.
Here and it what follows we use the  Bachmann-Landau ``small-$o$'' notation: $o\big(1/N^\alpha\big)$ is a function that (uniformly) vanishes in the thermodynamic limit faster than $1/N^\alpha$.  We further assume that $\delta_N(\rho,m,g)$ is a ``smooth'' function of the discrete variable $N$ in the following sense:
\begin{equation}\label{boundary scaling 2}
|\delta_N(\rho,m,g)-\delta_{N-1}(\rho,m,g)|=o(1/N).
\end{equation}

The fact that $E_N(V,m,g)$ scales as $N$ in the thermodynamic limit is a prerequisite for the thermodynamic description of the system. It can be rigorously proven, in particular, for fermions and fermion-boson mixtures in a broad range of physical conditions \cite{Fisher,Dyson,Lenard,Lieb}. It is also expected to hold for interacting bosons with short-range repulsion, although we are not aware of explicit proofs in this case (except for $D=1$, where bosons can be mapped to fermions~\cite{Girardeau_1960}). The thermodynamic scaling of $E_N(V,m,g)$ implies certain conditions on the short- and large-distance behavior of the potential $U({\bm r})$ \cite{ruelle1999statistical, gallavotti2013statistical}. In particular, $|U({\bm r})|$ should vanish faster than $|{\bm r}|^{-D}$ at large distances, otherwise the energy $E_N(V,m,g)/N$ can diverge to $+\infty$ in the thermodynamic limit \cite{ruelle1999statistical, gallavotti2013statistical,landau2013quantum}. One should also avoid the opposite peril -- the divergence of $E_N(V,m,g)/N$  to $-\infty$ in the thermodynamic limit, that shows up, in particular, for bosons with everywhere attractive potentials. Finally, for attractive potentials singular at ${\bm r}=0$ one should ensure that $U({\bm r})$ diverges slower that $1/|{\bm r}|^2$, otherwise the spectrum of the Hamiltonian can be unbounded from below already for $N=2$ \cite{landau2013quantum}.


While the exact scaling of the subleading boundary term with $N$ is not important for our argument, we mention that on physical grounds we expect $\delta_N(\rho,m,g) \sim N^{-1/D}$, $\partial_\rho\delta_N(\rho,m,g) \sim N^{-1/D}$,  $\partial_g\delta_N(\rho,m,g) \sim N^{-1/D}$ and $|\delta_N(\rho,m,g)-\delta_{N-1}(\rho,m,g)| \sim N^{-1-1/D}$, consistent with our assumptions \eqref{boundary scaling 1}, \eqref{boundary scaling 2}.

As is evident from eqs. \eqref{H homogeneous}-\eqref{varepsilon}, a trivial rescaling property holds for the energy per particle:
\begin{equation}\label{rescaling}
\varepsilon(\rho,m/c,c \, g)=c\, \varepsilon(\rho,m,g)
\end{equation}
for an arbitrary $c>0$. Taking the derivative of both sides of this identity with respect to  $c$ at $c=1$, one obtains


\bigskip

\noindent
{\it Lemma 1.}
\begin{equation}\label{identity}
\varepsilon=g\,\partial_g \varepsilon - m \, \partial_m \varepsilon
\end{equation}
Here and in what follows we  omit the arguments of  $\varepsilon(\rho,m,g)$ when this does not lead to a confusion.

\medskip

The main technical tool we are going to use is the lower bound on the ground state energy of an $N$-particle system in terms of ground state energies of a different $(N-1)$-particle system. A method for obtaining such bounds by partitioning an $N$-body Hamiltonian into sum of $(N-1)$-body Hamiltonians follows the idea by Anderson \cite{Anderson}. This idea was later developed in various directions, see e.g. \cite{Fisher,Dyson,Levy-Leblond,Hall_1967, Ader_1982,Nussinov_1984,Richard_1984,Khare_2001,uskov2019variational,Richard_2020}. We will use the following  bound~\cite{Khare_2001}:

\bigskip

 \noindent
{\it Lemma 2.}
\begin{equation}\label{Anderson bound 1}
E_N(V,m,g)\geq  \frac{N}{N-1}\,E_{N-1}\left( V,m,\frac{N-1}{N-2} \, g\right).
\end{equation}
\noindent
{\it Proof.} To make the paper self-contained, we reproduce the proof. The Hamiltonian \eqref{H homogeneous} is partitioned as
$H=\sum_{n=1}^N H_n$, where
\begin{equation}\label{Hn}
H_n= \frac1{N-1} \sum_{1\leq i \leq N \atop i\neq n} \, \frac{\bm{p}_i^2}{2m} + g\, \frac1{N-2}  \sum_{1\leq i<j\leq N \atop i,j\neq n} \, U(\bm{r}_i-\bm{r}_j).
\end{equation}
Note that the  terms with $i=n$ or $j=n$ are excluded from $H_n$. This means that while $H_n$ is a Hamiltonian of a system of $N$ particles (with the corresponding $n$-particle Hilbert space), the $n$'th particle is a ``spectator'' that is decoupled from the rest of the particles and does not contribute to energy. Since the infimum of a sum is not less than the sum of the infima, we get
\begin{equation}\label{derivation}
E_N(V,m,g) \geq \sum_{n=1}^N \inf_{\Psi:\, \|\Psi\|=1} \langle \Psi | H_n |  \Psi \rangle = N\, E_{N-1}\left(V,(N-1)m,\frac{g}{N-2}\right).
\end{equation}
Taking into account the rescaling property \eqref{rescaling}, one then obtains eq. \eqref{Anderson bound 1}.
 ~\qed

\subsection{Inequalities}

\medskip

The main technical result of the present section is contained in the following Lemma.

\vspace*{1cm}

\noindent
{\it Lemma 3.} 
\begin{equation}\label{main result 1}
g\,\partial_g \varepsilon  \leq\,  \rho\,\partial_\rho \varepsilon
\end{equation}

\vspace*{1cm}

\noindent
{\it Proof.} Eqs.  \eqref{varepsilon} and \eqref{Anderson bound 1} entail
\begin{align}\label{proof 1}
\varepsilon(\rho,m,g) \geq & \varepsilon\left(\frac{N-1}{N}\rho,m,\frac{N-1}{N-2}\,g\right) \nonumber \\
&+\left[\delta_{N-1}\left(\frac{N-1}{N}\rho,m,\frac{N-1}{N-2}\,g\right)-\delta_N (\rho,m,g)\right].
\end{align}
Eqs. \eqref{boundary scaling 1} and \eqref{boundary scaling 2} imply that  the term in square brackets  is $o(N^{-1})$. Expanding  the first term in the right hand side of eq. \eqref{proof 1} in powers of $N^{-1}$, we get
\begin{equation}\label{proof 2}
0 \geq N^{-1}\big( -\rho\,\partial_\rho\varepsilon(\rho,m,g)+g\,\partial_g\varepsilon(\rho,m,g)\big)+o(N^{-1}).
\end{equation}
In the thermodynamic limit this entails the  inequality  \eqref{main result 1}.
 ~\qed

 \bigskip

A remark regarding the inequality \eqref{Anderson bound 1} is in order. This inequality can be improved by accounting for the translation invariance of the system, with the result known as the Hall-Post inequality~\cite{Hall_1967,Khare_2001,Richard_2020}. This improvement considerably strengthens the bound for few-body systems \cite{Richard_2020}.  However, it does not lead to better bounds in the thermodynamic limit, therefore we do not employ it.

Combining the identity \eqref{identity} with the inequality \eqref{main result 1}, one immediately arrives at

 \vspace*{1cm}

\noindent
{\it Lemma 4.} 
\begin{equation}\label{corollary}
 - \rho\,\partial_\rho \varepsilon- m \, \partial_m \varepsilon+ 2 g\,\partial_g \varepsilon \leq\,  \varepsilon \leq\,  \rho\,\partial_\rho \varepsilon - m \, \partial_m \varepsilon
\end{equation}

\vspace*{1cm}

Note that the derivative $\partial_m \varepsilon$ is always non-positive. This follows from the Hellmann-Feynman theorem:  $\partial_m \varepsilon = \langle {\rm gs}| \partial_m H | {\rm gs}\rangle=- \langle {\rm gs}|\sum_{i=1}^{N} {\bm{p}_i^2}/({2m^2})| {\rm gs}\rangle\leq0$, where $ | {\rm gs}\rangle$ is the ground state.

\smallskip

We would like to reformulate Lemma 3 in more straightforward thermodynamic terms. To this end we introduce
the pressure
\begin{equation}
P=-\frac{\partial }{\partial V} \, E_N(V,m,g)
\end{equation}
and the interaction energy in the ground state
\begin{equation}\label{Eint}
E_{\rm int} = \langle {\rm gs}|\,  g \sum_{1\leq i<j \leq N} U(\bm{r}_i-\bm{r}_j) \, | {\rm gs}\rangle.
\end{equation}
By the Hellmann-Feynman theorem one gets $E_{\rm int} = N \,g\, \partial_g\varepsilon$. Taking this into account, we obtain from eq. \eqref{main result 1} the following

\vspace*{1cm}

\noindent
{\it Theorem 1.} 
\begin{equation}\label{main result Theorem}
E_{\rm int} \leq\,  P\,V.
\end{equation}

\vspace*{1cm}

\noindent
This is the main result of the present section.

\medskip

\subsection{Illustrations}

\begin{figure}[t] 
		\centering
\includegraphics[width=0.45\linewidth]{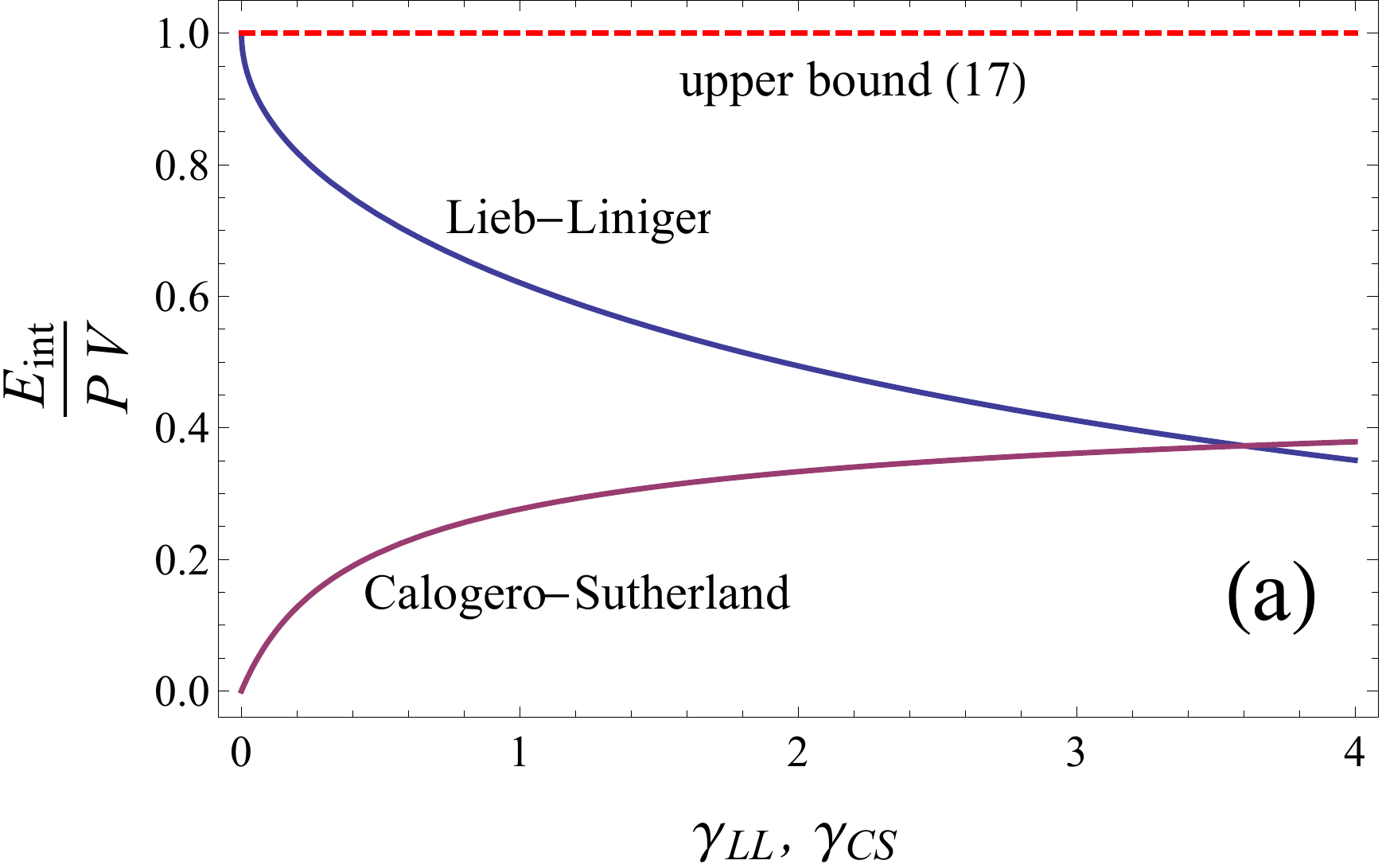}
~~~~~~
\includegraphics[width=0.45\linewidth]{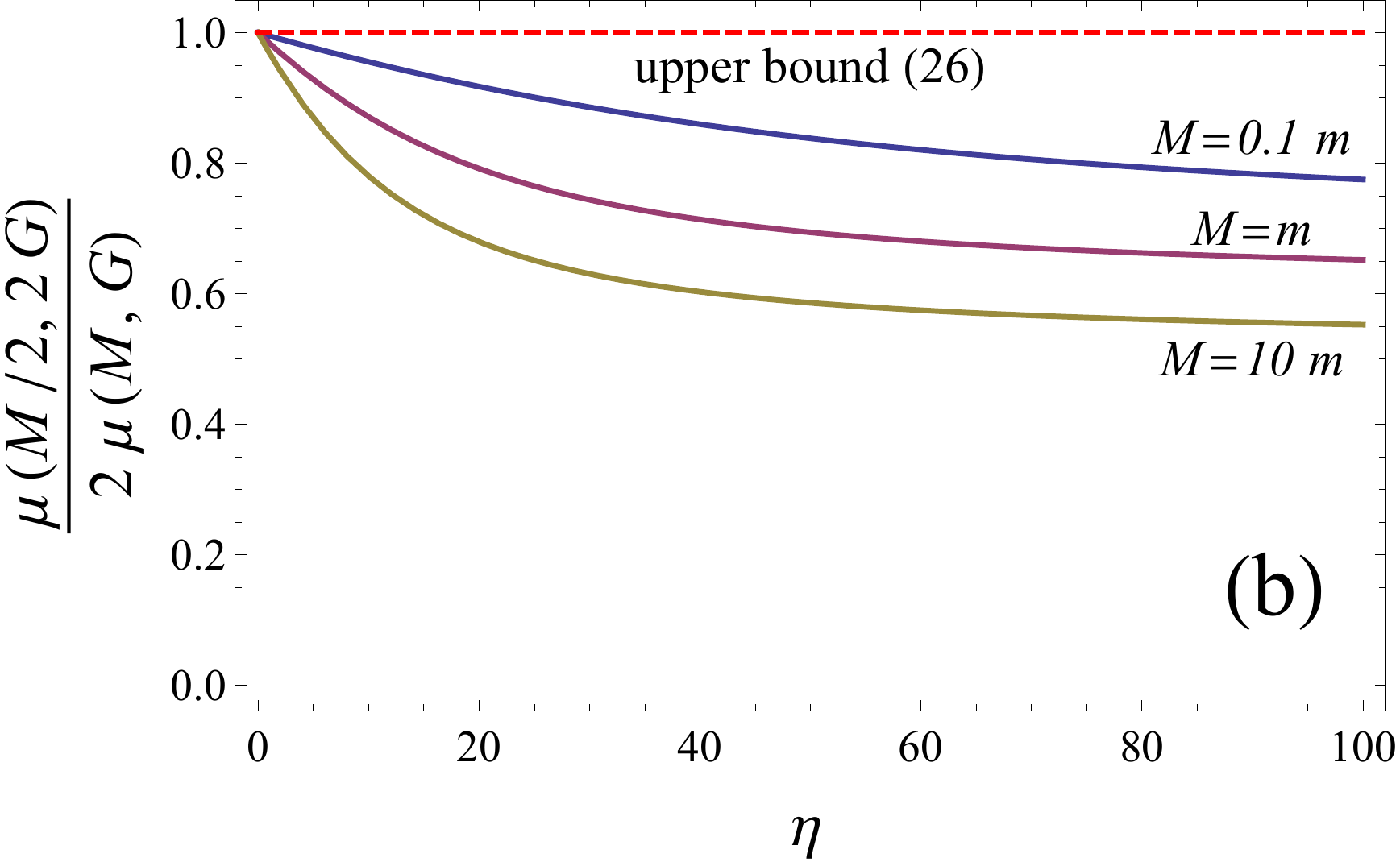}
        \caption{(a) Performance of the inequality \eqref{main result Theorem} as a function of coupling strength for the Lieb-Liniger and Calogero-Sutherland models. (b) Performance of the inequality \eqref{corollary chemical potentials} as a function of the relative coupling strength $\eta$ for impurities in a one-dimensional weakly-interacting Bose gas with contact interactions. The plots for three different mass ratios $M/m$ are shown. The healing length of the Bose gas is equal to $10/\rho$. }
		\label{fig}
\end{figure}

\subsubsection{Bosons.}


First consider bosons in the weak-coupling limit. The leading order of the  mean-field approximation gives $\varepsilon=(1/2) \rho \, g \int d {\bm r} \,U(\bm{r})$ (we assume that this quantity is positive). One can immediately verify that in this limit inequalities \eqref{main result 1}, \eqref{corollary}, \eqref{main result Theorem} saturate.

To illustrate the performance of the inequalities beyond the weak-coupling limit, we resort to the exactly solvable  Lieb-Liniger model of repulsive one-dimensional bosons with the contact interaction $U(x)=\delta(x)$  and $g>0$ \cite{Lieb_Liniger_1963}. The ground state energy as a function of the dimensionless coupling $\gamma_{LL}=m g/\rho$ can be found by solving a system of coupled integral equations  \cite{Lieb_Liniger_1963}.  An approximation to this solution, which is extremely accurate up to $\gamma=15$, has been obtained recently in \cite{Lang_2017_LL_gs_energy}. Using this approximation, we plot the ratio $E_{\rm int} /  (P\,V)$ in Fig. \ref{fig} (a). The inequality \eqref{main result Theorem} bounds this ratio from above by one. One can see that at larger couplings the inequality becomes less strong, but retains 50\% accuracy up to moderate $\gamma\simeq 2$.

\subsubsection{Fermions.}


For fermions the inequalities are, in general, quite loose. This is particularly manifest for the free-fermion case, where $E_{\rm int}=0$ while $P\,V$ is finite.

To illustrate the interacting case, we resort to the exactly solvable Calogero-Sutherland model \cite{Calogero_1969}\cite{Sutherland_1971}. This is a one-dimensional model with the potential $U(x)=1/x^2$, and we assume $g>0$.   The ground state energy per particle as a function of the dimensionless coupling $\gamma_{CS}=g m$ is given by
$
\varepsilon=({\pi^2}/12) \,(\rho^2/m)\,\big(1+2\gamma_{CS}+\sqrt{1+4\gamma_{CS}}\,\big)
$~\cite{Sutherland_1971}.
We plot the ratio $E_{\rm int} /  (P\,V)$ in Fig.~\ref{fig}~(a). This ratio grows from zero at the vanishing coupling to $1/2$ in the limit of the infinitely strong coupling.

\section{System with mobile impurities}

\medskip

\subsection{Inequalities}

\medskip

Now we turn to the case when a few impurity particles, i.e. particles of a different type, are added to the system. Let us start from the case of a single impurity particle. The corresponding Hamiltonian reads
\begin{equation}\label{H1}
H_1=H + \frac{\bm{P}^2}{2M} + G \sum_{i=1}^{N}  \tilde U(\bm{R}-\bm{r}_i),
\end{equation}
where $H$ is the Hamiltonian \eqref{H homogeneous} of the majority species, $\bm{P}$ and $\bm{R}$ are, respectively, the operators of the momentum and the coordinate of the impurity, $M>0$ is its  mass, and  $G \, \tilde U(\bm{R}-\bm{r}_j)$ is an interaction potential between the impurity and the majority species. Here, again, we single out a multiplicative coupling constant $G>0$.

The ground state energy of the Hamiltonian \eqref{H1} is denoted by $E_{N,1}(V,m,g,M,G)$. We assume that it can be represented as
\begin{equation}\label{chemical potential}
E_{N,1}(V,m,g,M,G)=E_N(V,m,g)+\mu(\rho,m,g,M,G)+o(1)
\end{equation}
in the thermodynamic limit $N,V\rightarrow\infty$, $N/V=\rho$. Here $\mu$ is the chemical potential of the impurity particle.

Next we consider a system with two identical impurities. Its Hamiltonian reads
\begin{equation}\label{H2}
H_2=H + \frac{\bm{P}_1^2}{2M}+\frac{\bm{P}_2^2}{2M} + G \sum_{i=1}^{N}  \big(\tilde U(\bm{R}_1-\bm{r}_i)+\tilde U(\bm{R}_2-\bm{r}_i)\big),
\end{equation}
where $\bm{P}_1$, $\bm{P}_2$ are momenta and $\bm{R}_1$, $\bm{R}_2$ are coordinates of the two impurities. Note that we assume that there is no direct interaction between the impurities.

We assume that the ground state energy $E_{N,2}(V,m,g,M,G)$ of the Hamiltonian \eqref{H2} can be expanded as
\begin{equation}\label{binding energy}
E_{N,2}(V,m,g,M,G)=E_N(V,m,g)+2\mu(\rho,m,g,M,G)-\epsilon_b(\rho,m,g,M,G)+o(1),
\end{equation}
where $\epsilon_b (\rho,m,g,M,G) \geq 0$ is the binding energy. The rationale behind this assumption is as follows.  If two impurities do not form a bound state, they  reside far apart from each other (an average distance between the two being on the order of the system size). Then adding a second impurity to the system simply increases the energy~\eqref{chemical potential} by the chemical potential, thus  $\epsilon_b (\rho,m,g,M,G) = 0$ in this case. Alternatively, the two impurities can form a bound state when immersed in the bulk of the background species, despite the absence of the direct interaction between them. This effect is a precursor of the phase separation which can take place e.g. in repulsive mixtures of Fermi gases \cite{Cui_2013}. In this case the energy is lowered with respect to the energy of distant impurities, which implies $\epsilon_b (\rho,m,g,M,G) > 0$.

In order to derive inequalities on the chemical potential and binding energy, we need an analog of Lemma 2 adapted for two species:

\bigskip

 \noindent
{\it Lemma 3.}
\begin{equation}\label{Anderson bound 2}
E_{N,2}(V,m,g,M,G) \geq  E_{N,1}(V,m,g,M/2,2G).
\end{equation}
\noindent
{\it Proof.} Analogously to the proof of Lemma 2, we use the method by Anderson \cite{Anderson}. This time, however, we  partition the Hamiltonian $H_2$ in {\it two} parts:
\begin{equation}\label{H2=H1+H1}
H_2= \big(\frac12 H+\frac{\bm{P}_1^2}{2M} + G \sum_{i=1}^{N}  \tilde U(\bm{R}_1-\bm{r}_i)\big)+
 \big(\frac12 H+\frac{\bm{P}_2^2}{2M} + G \sum_{i=1}^{N}  \tilde U(\bm{R}_2-\bm{r}_i)\big).
\end{equation}
This implies
\begin{equation}\label{Anderson bound 2}
E_{N,2}(V,m,g,M,G) \geq 2 \, E_{N,1}(V,2m,g/2,M,G)=E_{N,1}(V,m,g,M/2,2G),
\end{equation}
where the latter identity follows from the generalization of the rescaling property \eqref{rescaling} to a two-component system.
 ~\qed

\medskip

Lemma 3 along with eqs. \eqref{chemical potential}, \eqref{binding energy} imply

\vspace*{1cm}

\noindent
{\it Theorem 2.} 

\medskip

\begin{equation}\label{main result 2}
 \epsilon_b (\rho,m,g,M,G) \leq 2\, \mu(\rho,m,g,M,G)-\mu(\rho,m,g,M/2,2G).
\end{equation}

\vspace*{1cm}

This theorem bounds from above the binding energy of two impurities immersed in medium. This bound parallels the results of \cite{Ader_1982,Nussinov_1984,Richard_1984,Khare_2001,uskov2019variational,Richard_2020}  obtained for binding energies of few-body systems in vacuum.

Assuming further that $\epsilon_b \geq 0$, as explained above, we arrive at 

\vspace*{1cm}

\noindent
{\it Corollary.} 

\medskip

\begin{equation}\label{corollary chemical potentials}
\mu(\rho,m,g,M/2,2G)\leq 2\, \mu(\rho,m,g,M,G).
\end{equation}

\vspace*{1cm}

This corollary relates the chemical potentials of impurities with different masses and couplings. 

\subsection{Illustrations}

\medskip

The inequality \eqref{main result 2} saturates in the limit of large $M/m$ (whatever other parameters of the system are) where  the impurities can be considered as static potentials \cite{petkovic2021mediated}.

In the limit of weak coupling between the impurities and the majority species one straightforwardly obtains $\mu(\rho,m,g,M,G) = \rho \, G \int d {\bm r} \,\tilde U(\bm{r}) +o(G)$. This expression saturates the inequality \eqref{corollary chemical potentials}. It also implies that, according to the bound \eqref{main result 2}, $\epsilon_b (\rho,m,g,M,G) =o(G)$. This is indeed the case: in two and three dimensions bound states do not exist for $G$ below a critical one (i.e. $\epsilon_b (\rho,m,g,M,G) =0$) \cite{Camacho-Guardian_2018}; in one dimension  $\epsilon_b (\rho,m,g,M,G) =O(G^2)$ \cite{Naidon_2018}.

The performance of the inequality \eqref{corollary chemical potentials} can be illustrated beyond these limits for an impurity immersed in a weakly interacting one-dimensional Bose gas with contact interactions $U(x)=\tilde U(x)=\delta(x)$. There an approximate expression is available that is very accurate in the whole range of couplings $G>0$ \cite{Jager_2020}, as independently confirmed by diffusion Monte Carlo simulations~\cite{Grusdt_2017}. We plot the ratio of the left hand side to the right hand side of eq. \eqref{corollary chemical potentials} as a function of the relative coupling strength $\eta=G/g$ in Fig. \ref{fig} (b). One can see that the inequality remains reasonably tight for a wide range of parameters, particularly for light impurities.

\medskip

\section{Summary and outlook}

\medskip

For a single-component system, we have derived the bound \eqref{main result Theorem} (that can be rewritten in different forms \eqref{main result 1}, \eqref{corollary}). For bosons, this bound is saturated in the weak-coupling limit. We have tested its performance for arbitrary couplings in  the exactly solvable Lieb-Liniger model, see Fig. \ref{fig} (a). For fermions, the performance of the bound is, in general,  much worse, as can be illustrated by the exactly-solvable Calogero-Sutherland model, see Fig. \ref{fig} (a).

We have also considered a system with one or two impurity particles added. We assume that in vacuum the impurities  do not interact one with another; nevertheless, in the medium an effective interaction between the impurities can be induced. We have derived the inequality \eqref{main result 2} that relates chemical potentials of impurities with the binding energy of the in-medium bound state they may form. This inequality is saturated in the limit of the weak impurity-medium interactions, as well as for infinitely-heavy impurities. Under an additional, physically natural assumption of non-negativity of the binding energy, the inequality \eqref{main result 2} entails the inequality \eqref{corollary chemical potentials} that bounds chemical potentials alone. We have verified that for  a particular medium --  the one-dimensional weakly-interacting Bose gas --  this latter bound is reasonably strong for a wide range of impurity-medium couplings, see Fig.~\ref{fig} (b).

We anticipate that our method can be generalized to more sophisticated situations, e.g. mixtures containing multiple species. 

\medskip
\section*{Acknowledgements.} The work was supported by the Russian Foundation for Basic Research under the grant No. 18-32-20218.

\section*{References}
\bibliographystyle{iopart-num}
\bibliography{inequalities}

\end{document}